\documentclass[aps,prl,preprint,tightenlines,superscriptaddress,showpacs,byrevtex]{revtex4}
\usepackage{graphicx} 
\usepackage{dcolumn}  

\graphicspath{{ps}}

\newcommand{\ee}{e^{+} e^{-}}
\newcommand{\bbar}{B\bar{B}}
\newcommand{\leplep}{\ell^{+}\ell^{-}}
\newcommand{\jp}{J/\psi}
\newcommand{\psip}{\psi '}

\newcommand{\mumu}{\mu^{+}\mu^{-}}
\newcommand{\pipi}{\pi^{+}\pi^{-}}

\newcommand{\kstr}{K^{*}(890)}
\newcommand{\ktwostr}{K_2^{*}(1430)}

\newcommand{\Mbc}{M_{\rm bc}}
\newcommand{\DE}{\Delta E}

\newcommand{\rt}{\rightarrow}
\newcommand{\etal}{\em et al.}

\begin{document}


\preprint{\vbox{ \hbox{   }
}}

\title{ \quad\\[1.0cm]   
Observation of a resonance-like structure 
in the $\pi^{\pm}\psip$ mass distribution
in exclusive 
$B \rt K \pi^{\pm}\psip$ decays
}

\affiliation{Budker Institute of Nuclear Physics, Novosibirsk}
\affiliation{Chiba University, Chiba}
\affiliation{University of Cincinnati, Cincinnati, Ohio 45221}
\affiliation{Justus-Liebig-Universit\"at Gie\ss{}en, Gie\ss{}en}
\affiliation{The Graduate University for Advanced Studies, Hayama}
\affiliation{Gyeongsang National University, Chinju}
\affiliation{Hanyang University, Seoul}
\affiliation{University of Hawaii, Honolulu, Hawaii 96822}
\affiliation{High Energy Accelerator Research Organization (KEK), Tsukuba}
\affiliation{Institute of High Energy Physics, Chinese Academy of Sciences, Beijing}
\affiliation{Institute of High Energy Physics, Vienna}
\affiliation{Institute of High Energy Physics, Protvino}
\affiliation{Institute for Theoretical and Experimental Physics, Moscow}
\affiliation{J. Stefan Institute, Ljubljana}
\affiliation{Kanagawa University, Yokohama}
\affiliation{Korea University, Seoul}
\affiliation{Kyungpook National University, Taegu}
\affiliation{\'Ecole Polytechnique F\'ed\'erale de Lausanne (EPFL), Lausanne}
\affiliation{University of Ljubljana, Ljubljana}
\affiliation{University of Maribor, Maribor}
\affiliation{University of Melbourne, School of Physics, Victoria 3010}
\affiliation{Nagoya University, Nagoya}
\affiliation{Nara Women's University, Nara}
\affiliation{National Central University, Chung-li}
\affiliation{National United University, Miao Li}
\affiliation{Department of Physics, National Taiwan University, Taipei}
\affiliation{H. Niewodniczanski Institute of Nuclear Physics, Krakow}
\affiliation{Nippon Dental University, Niigata}
\affiliation{Niigata University, Niigata}
\affiliation{University of Nova Gorica, Nova Gorica}
\affiliation{Osaka City University, Osaka}
\affiliation{Osaka University, Osaka}
\affiliation{Panjab University, Chandigarh}
\affiliation{Saga University, Saga}
\affiliation{University of Science and Technology of China, Hefei}
\affiliation{Seoul National University, Seoul}
\affiliation{Sungkyunkwan University, Suwon}
\affiliation{University of Sydney, Sydney, New South Wales}
\affiliation{Tata Institute of Fundamental Research, Mumbai}
\affiliation{Toho University, Funabashi}
\affiliation{Tohoku Gakuin University, Tagajo}
\affiliation{Department of Physics, University of Tokyo, Tokyo}
\affiliation{Tokyo Institute of Technology, Tokyo}
\affiliation{Tokyo Metropolitan University, Tokyo}
\affiliation{Tokyo University of Agriculture and Technology, Tokyo}
\affiliation{Virginia Polytechnic Institute and State University, Blacksburg, Virginia 24061}
\affiliation{Yonsei University, Seoul}
  \author{S.-K.~Choi}\affiliation{Gyeongsang National University, Chinju} 
  \author{S.~L.~Olsen}\affiliation{University of Hawaii, Honolulu, Hawaii 96822}\affiliation{Institute of High Energy Physics, Chinese Academy of Sciences, Beijing} 
   \author{I.~Adachi}\affiliation{High Energy Accelerator Research Organization (KEK), Tsukuba} 
   \author{H.~Aihara}\affiliation{Department of Physics, University of Tokyo, Tokyo} 
   \author{V.~Aulchenko}\affiliation{Budker Institute of Nuclear Physics, Novosibirsk} 
   \author{T.~Aushev}\affiliation{\'Ecole Polytechnique F\'ed\'erale de Lausanne (EPFL), Lausanne}\affiliation{Institute for Theoretical and Experimental Physics, Moscow} 
   \author{T.~Aziz}\affiliation{Tata Institute of Fundamental Research, Mumbai} 
   \author{A.~M.~Bakich}\affiliation{University of Sydney, Sydney, New South Wales} 
   \author{V.~Balagura}\affiliation{Institute for Theoretical and Experimental Physics, Moscow} 
   \author{I.~Bedny}\affiliation{Budker Institute of Nuclear Physics, Novosibirsk} 
% ...
\collaboration{The Belle Collaboration}
\begin{abstract}
A distinct peak is
observed in the $\pi^{\pm}\psip$ invariant mass distribution  
near $4.43$~GeV in  $B \rightarrow K \pi^{\pm} \psip$ decays.   
A fit using a
Breit-Wigner resonance shape yields a peak mass and width of
$M=4433\pm 4 {\rm ~(stat)} \pm 2 {\rm ~(syst)}$~MeV 
and $\Gamma = 45 ^{+18}_{-13}{\rm ~(stat)} ^{+30} _{-13} {\rm 
~(syst)}$~MeV.  The product branching fraction is determined to be 
${\mathcal B}(B^0\rt K^{\mp} Z^{\pm}(4430))\times {\mathcal 
B}(Z^{\pm}(4430)\rt 
\pi^{\pm}\psip) = (4.1 \pm 1.0 {\rm (stat)}\pm 1.4 {\rm (syst)})\times 
10^{-5}$, where $Z^{\pm}(4430)$ is used to denote
the observed structure.  
The statistical significance of the observed peak
is  $6.5\sigma$.
These results are obtained from a $605\,{\rm fb}^{-1}$ data sample 
that contains 657 million $B\bar{B}$ pairs collected
near the $\Upsilon(4S)$ resonance
with the Belle detector at the KEKB asymmetric energy $e^+ e^-$
collider.

\end{abstract}


\maketitle

\tighten

{\renewcommand{\thefootnote}{\fnsymbol{footnote}}}
\setcounter{footnote}{0}


An unexpected bonus from the $B$-factory experiments has been the 
discovery of a large number of charmonium-like meson 
states~\cite{skchoi_etac2s,uehara_chic2,skchoi_x3872,aubert_y4260,
kabe_x3940,skchoi_y3940,aubert_y4325,yuan_y4096,yuan_y4660}.  
While two of these have been identified as likely
candidates for ``missing'' charmonium
states, namely the $\eta_c'$~\cite{skchoi_etac2s}  and the 
$\chi_{c2}'$~\cite{uehara_chic2},  others have properties
that are at odds with expectations of the charmonium model. These
latter include the $X(3872)$~\cite{skchoi_x3872} and  
$Y(4260)$~\cite{aubert_y4260}, which are seen to decay to $\pipi\jp$;
the $X(3940)$~\cite{kabe_x3940}, seen in $D^*\bar{D}$; the
$Y(3940)$, seen in $\omega\jp$~\cite{skchoi_y3940};
and the $Y(4325)$~\cite{aubert_y4325}, seen in $\pipi\psip$.  
Recently Belle
reported a second $\pipi\jp$ mass enhancement below the 
$Y(4260)$~\cite{yuan_y4096} and has shown that the
$Y(4325)\rt\pipi\psip$ signal,  reported by BaBar to have a width of 
$\Gamma=172\pm33$~MeV, is better fitted with two narrower
peaks, one at 4361~MeV with $\Gamma = 70\pm 20$~MeV
and a second at 4664~MeV, with $\Gamma = 40\pm 
17$~MeV~\cite{yuan_y4660}.

Proposed assignments for these states have included:
multiquark states, either of the $(c\bar{q},\bar{c}q)$ ``molecular''
type~\cite{molecular} or $[cq,\bar{c}\bar{q}]$ diquark-antidiquark
type~\cite{diquark} (here $c$ represents a charmed quark and $q$
either a $u$-, $d$- or $s$-quark); hybrid $c\bar{c}$-gluon 
mesons~\cite{hybrid};
or other missing charmonium states where the masses 
predicted by potential models are 
drastically modified by nearby $D^{(*)}\bar{D^{(*)}}$ 
thresholds~\cite{eichten, chao}.
A characteristic that clearly distinguishes multiquark 
states from hybrids or charmonia is the
possibility to have charmonium-like mesons with non-zero
charge (e.g. $[cu\bar{c}\bar{d}]$), strangeness
($[cd\bar{c}\bar{s}]$)  or both ($[cu\bar{c}\bar{s}]$)~\cite{chiu}.
These considerations motivated a search for charmonium-like 
mesons with non-zero electric charge.

Here we report the observation of a relatively narrow peak
in the $\pi^+\psip$ invariant mass distribution 
produced in exclusive  $B\rt K\pi^+\psip$ decays~\cite{conj}.
The results are based on an analysis of a 657~million 
$\bbar$ event sample collected in the Belle detector 
operating at the KEKB asymmetric energy $e^+e^-$ collider. 
The data were  accumulated  at a  center-of-mass system 
(cms) energy of $\sqrt{s} = 10.58$~GeV, corresponding to 
the mass of the $\Upsilon(4S)$ resonance.  
KEKB is described in detail in  Ref.~\cite{KEKB}.

The Belle detector, described in Ref.~\cite{Belle}, 
is a large-solid-angle magnetic
spectrometer that
consists of a silicon vertex detector,
a 50-layer cylindrical drift chamber, an array of
aerogel threshold Cherenkov counters,  
a barrel-like arrangement of time-of-flight 
scintillation counters, and an electromagnetic calorimeter
comprised of CsI(Tl) crystals  located inside
a superconducting solenoid coil that provides a 1.5~T
magnetic field.  An iron flux-return located outside of 
the coil is instrumented to detect $K^0_L$ mesons and to identify
muons.


We select events of the type $B \rt K \pi^+\psip$,
where the $\psip$ decays either to $\leplep$ or 
$\pi^+\pi^-\jp$ with $\jp\rt\leplep$ ($\ell = e$ or $\mu$).
Both charged and neutral ($K^0_S\rt\pi^+\pi^-$) kaons 
are used.  Charged tracks other than $K^0_S\rt\pi^+\pi^-$
secondaries are required to originate 
from the beam-beam interaction point. 
The charged kaon, lepton and pion selection requirements 
are described in Ref.~\cite{skchoi_x3872}; those for
neutral kaons are described in Ref.~\cite{skchoi_y3940}.

For $\psip (\jp)\rt\leplep$ candidates we require
the invariant mass of the lepton pair to be within 
20~MeV of the ${\psip}(\jp)$ mass. 
For $\psip (\jp)\rt\ee$ candidates, we include 
photons that
are within 50~mrad of the $e^+$ or $e^-$ tracks in 
the invariant mass calculation.  For
$\psip\rt\pipi\jp$ candidates, we require the $\pipi$
invariant mass to be greater than 0.44~GeV and 
$|M(\pipi\leplep) - M(\leplep)-0.589~{\rm GeV}|<0.0076$~GeV,
which is $\pm 2.5\sigma$, where $\sigma$ is the rms
resolution.

We suppress continuum $\ee \rt q\bar{q}$ events,
where $q=u,d,s$ or $c$, by requiring $R_2 < 0.4$, 
where $R_2$ is the
second normalized
Fox-Wolfram event-shape moment~\cite{SFW}. 
We also require $|\cos\theta_B|<0.9$, where
$\theta_B$ is the angle between the $B$ meson
and $e^+$ beam directions~\cite{thetaB}.

We identify $B$ mesons using the beam-constrained mass 
$\Mbc=\sqrt{E_{\rm beam}^2 - p_B^2}$ and the energy difference
$\Delta E = E_{\rm beam} - E_B$, where $E_{\rm beam}$ is the
cms beam energy, $p_B$ is the vector sum of the 
cms momenta of the $B$ meson decay products and $E_B$ is their
cms energy sum.    We select events with 
$|\Mbc -m_B| < 0.0071$~GeV ($m_B=5.279$~GeV, is the
world average $B$-meson mass~\cite{PDG})
and $|\DE |< $ 0.034 GeV, which are $\pm 2.5\sigma$ 
windows around the nominal peak values.

The invariant mass of the selected $B\rt K\pi\psip$ candidate 
tracks is kinematically constrained to equal $m_B$. 
This improves the $\psip\rt\leplep$ ($\jp\rt\leplep$)
mass resolution to  $\sigma=4.4$~MeV ($5.3$~MeV).   
We require  $M(\leplep)$  computed with
the fitted lepton four-vectors to be within $\pm 2.5\sigma$
of $m_{\psip}$ ($m_{\jp}$), the world average 
$\psip$ ($\jp$) mass~\cite{PDG}. 

For the $\psip\rt\leplep$ mode we compute
$M(\pi\psip)$ as  $M(\pi\leplep)-M(\leplep)
+m_{\psip}$; for $\psip\rt\pipi\jp$ decays,
we  use 
$M(\pi\psip)=M(\pi\pipi\jp)-M(\pipi\jp)+m_{\psip}$.
Simulations of the two $\psip$ decay modes indicate 
that the  experimental resolution for $M(\pi^+\psip)$ is 
$\sigma\simeq 2.5$~MeV for both modes.

\begin{figure}[htb]
\includegraphics[width=0.6\textwidth]{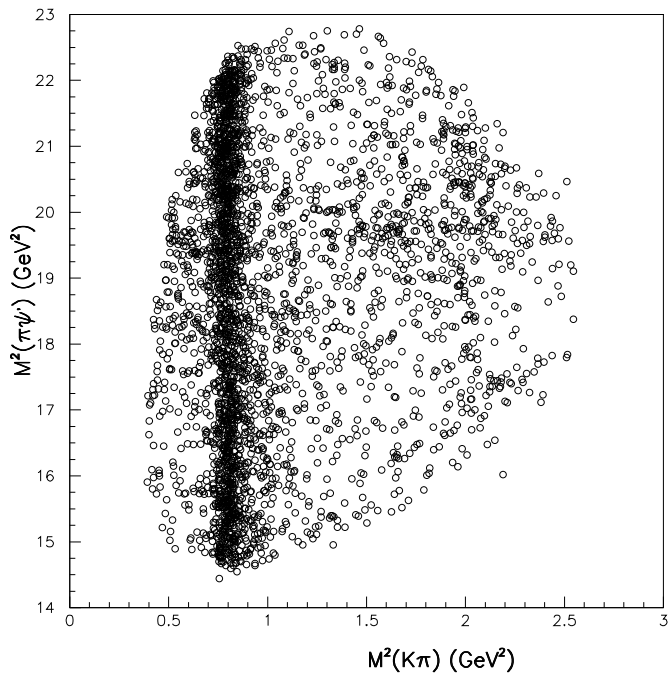}
\caption{
 The $M^2(K\pi)$ (horizontal) {\it vs.}
$M^2(\pi\psip)$ (vertical) Dalitz-plot distribution 
for  $B^0\rt K^-\pi^+\psip$ candidate events.
}
\label{fig:dalitz}
\end{figure}

Figure~\ref{fig:dalitz} shows a Dalitz plot of
$M^2(K\pi^+)$ (horizontal) $vs.$ $M^2(\pi^+\psip)$
(vertical) for the $B\rt K\pi^+\psip$ candidate events.  
Here, a distinct band at $M^2_{K\pi}\simeq 0.8$~GeV$^2$,
corresponding to $B\rt \kstr\psip$; $\kstr\rt K\pi$,
is evident.  
In addition, there are signs of a $\ktwostr$  signal
near $M^2_{K\pi}= 2.0$~GeV$^2$.
The $B\rt K^*(890)\psip$ events 
are used to calibrate
the $\Mbc$ and $\DE$ peak positions and widths.

Some clustering of events in a horizontal band 
is evident in  the upper half of the Dalitz plot
near $M^2(\pi\psip)\simeq 20$~GeV$^2$.  To study these
events with the effects of the known $K\pi$ resonant 
states minimized, we restrict
our analysis to the events with
$|M(K\pi) - m_{\kstr}|\ge 0.1$~GeV and
$|M(K\pi) - m_{\ktwostr}|\ge 0.1$~GeV. In the following, 
we refer to this requirement as the $K^*$ veto.

\begin{figure}[htb]
\includegraphics[width=0.6\textwidth]{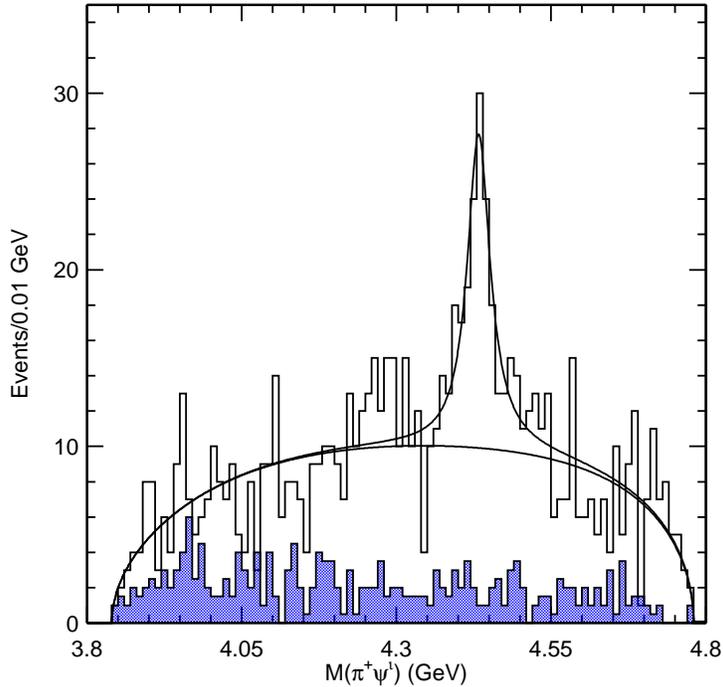}
\caption{
The $M(\pi^+\psip)$ distribution for 
events in the $\Mbc$-$\DE$ signal region
and with the $K^*$ veto applied.
The shaded histogram show the scaled results
from the $\DE$ sideband.  The solid curves show the
results of the fit described in the text.
}
\label{fig:mpipsip_prl}
\end{figure}

The open histogram in 
Fig.~\ref{fig:mpipsip_prl}
shows the $M(\pi^+\psip)$ distribution for selected events 
with the $K^*$ veto applied.  The bin width is  10~MeV. 
The  shaded histogram shows the scaled distribution
from $\DE$ sidebands ($|\DE \pm 0.070|< 0.034$~GeV).
Here a strong
enhancement is evident near $M(\pi\psip)\sim$4.43~GeV.

We perform a binned maximum-likelihood fit to the $M(\pi\psip)$ invariant 
mass distribution using a relativistic
$S$-wave Breit-Wigner (BW) function to model the peak plus
a smooth phase-space-like function $f_{\rm cont}(M)$, where 
$f_{\rm cont}(M) =
{\mathcal N_{\rm cont}}q^*(Q^{1/2} + A_1 Q^{3/2} + A_2 Q^{5/2}).$
Here $q^*$ is the momentum of the $\pi^+$ in the $\pi\psip$
rest frame and $Q=M_{\rm max}-M$, where $M_{\rm max}=4.78$~GeV is the 
maximum $M(\pi\psip)$ value possible for $B\rt K\pi\psip$
decay. The normalization ${\mathcal N_{\rm cont}}$ and two shape parameters
$A_1$ and $A_2$ are free parameters in the fit.  This 
form for $f_{\rm cont}(M)$ is chosen because it mimics two-body 
phase-space behavior at the lower and upper mass boundaries.
(Since the $M(\pi\psip)$ distribution for the non-peaking 
$B$-decay events and the $\DE$ sideband events have a similar shape, 
we represent them both with a single function.)

The results of the fit, shown as smooth curves in 
Fig.~\ref{fig:mpipsip_prl}, are tabulated in 
Table~\ref{tbl:fits_all}.  
The fit quality is $\chi^2 = 80.2$ for 
94~degrees of freedom.
The significance of the peak, determined from
the change in log likelihood when the signal and its associated
degrees of freedom are removed from the fit,
is $6.5\sigma$.

%
%
%

\begin{table}[htb]
\caption{Results of the fit shown in Fig.~\ref{fig:mpipsip_prl}.}
\label{tbl:fits_all}
\begin{tabular}
{@{\hspace{0.2cm}}c@{\hspace{0.2cm}}|@{\hspace{0.2cm}}c@{\hspace{0.2cm}}|
@{\hspace{0.2cm}}c@{\hspace{0.2cm}}|@{\hspace{0.2cm}}c@{\hspace{0.2cm}}}
\hline \hline
$N_{\rm sig}$    &  ${\mathcal N}_{\rm cont}$   
         &  BW Mass (GeV)       &  $\Gamma$~(GeV)    \\
\hline
 $121 \pm 30$ &  $766\pm 39 $  &
           $4.433 \pm0.004$ & $0.045^{+0.018}_{-0.013}$\\ 
\hline \hline
\end{tabular}
\end{table}




We fit $M(\pi\psip)$ distributions for various subsets of
the data.  The results are summarized in
Table~\ref{tbl:fits_separate}.

There are significant
({\it i.e.} significance of $4.5\sigma$ or more) 
signals with consistent 
mass values in both the $\psip\rt\pipi\jp$ and 
$\psip\rt\leplep$ subsamples.  However, 
the width of the peak
in the $\psip\rt\leplep$ subsample is substantially wider than
that for the $\pipi\jp$ subsample.   
Fitting the two measured widths to a common value gives a $\chi^2=4.8$
for 1 degree of freedom.  The corresponding confidence level is 
$\simeq 3\%$.

The fitted values for the signal yields are highly correlated with the
widths.  To compare the yields in each subchannel, we
refit the distributions using a width that is fixed at 
the $\Gamma=0.045$~GeV value determined from the common fit.
The ratio of ``constrained'' signal yields for the
$\pipi\jp$ and $\leplep$ subsamples (see Table~\ref{tbl:fits_separate})
is $1.09 \pm0.35$, in good agreement with the MC-determined
acceptance ratio of 1.23.


\begin{table}[htb]
\caption{Results of fits to different subsamples of the data.}
\label{tbl:fits_separate}
\begin{center}
\begin{tabular}{l||c|c|c|c}
\hline\hline
Subset &  Mass   &  Width 
   &  Signif. &  Constr. yield\\
        &     (GeV)           &    (GeV)
      &   ($\sigma$)                    &  ($\Gamma =0.045$GeV)                 
    
\\
\hline\hline
$\pipi\jp$   &  $4.435 \pm 0.004$ 
&  $0.026^{+0.013}_{-0.008}$  & 4.5  & $64 \pm 15$ \\
$\leplep$    &  $4.435\pm 0.010$
& $0.094^{+0.042}_{-0.030}$   &   4.7 & $59\pm 13$\\
\hline
$\ee$   &  $4.430\pm0.009$ 
&  $0.056^{+0.028}_{-0.020}$  & 3.5  & $41 \pm 12$ \\
$\mumu$ &  $4.434\pm0.004$ 
&  $0.038^{+0.023}_{-0.013}$  & 5.2  & $80 \pm 16$ \\
\hline
$K^{\pm}\pi^{\mp}\psip$  &  $4.434\pm0.005$ 
&  $0.048^{+0.019}_{-0.014}$  & 6.0  & $102 \pm 18$ \\
$K^0_S\pi^{\mp}\psip$       &  $4.430\pm0.009$ 
&  $0.048$-fixed  & 2.0  & $19 \pm 8$ \\\hline
$K^*$ veto   &  $4.437\pm0.005$ 
&  $0.063 ^{+0.024}_{-0.017}$  & 7.1  & $170 \pm 26$ \\
\hline\hline
\end{tabular} 
\end{center}
\end{table}

Table~\ref{tbl:fits_separate} also shows the results from dividing
the data sample into $\leplep = \ee $ and $\mumu$, and
charged kaon and $K^0_S\rt\pipi$ subsets. We see signals in
both the $\ee$ ($3.5\sigma$) and $\mumu$ modes ($5.2\sigma$)
with consistent mass and width values, and with constrained yields
that are consistent with the expected $\ee$/$\mumu$ acceptance
ratio of 0.61.  There are too few events in the $K^0_S$ sample
to enable a stable fit with yield, mass and width all allowed
to vary.  With the width fixed at the value found for the charged
kaon sample,  the fit returns a $19\pm8$ event signal with $2.0\sigma$ 
significance and a consistent mass value.  The observed signal 
yield in the $K^0_S$ sample agrees with expectations~\cite{ks_yield}
based on scaling
the charged kaon signal by the $K^0_S/K^{\pm}$ acceptance ratio
(0.19). 

The last row of Table~\ref{tbl:fits_separate} shows the results of
a fit to the $M(\pi\psip)$ distribution for the case
where the $K^*$ veto is replaced by a less stringent
requirement that only eliminates the core
of the $\kstr$ peak: $|M(K\pi) - m_{\kstr}|\ge 0.05$~GeV. Here 
the observed signal increases and its statistical significance improves to
$7.1\sigma$.

\begin{figure}[htb]
\includegraphics[width=0.6\textwidth]{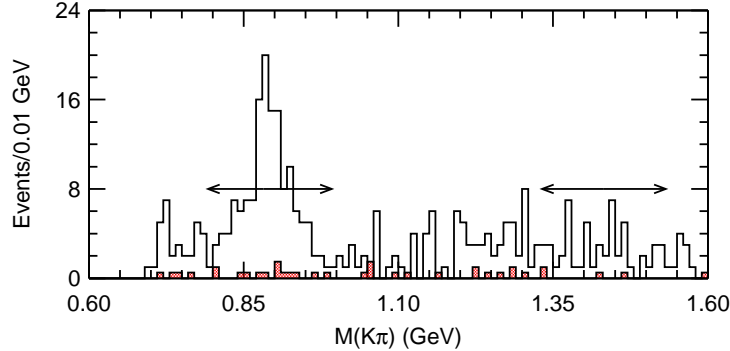}
\caption{ 
The $M(K\pi)$ distribution for events within $\pm 0.03$~GeV
of the 4.43~GeV peak.  Here the $K^*$ veto, which excludes
the regions indicated by double-sided arrows, has been 
removed.
The shaded histogram indicates the $\DE$-sideband-determined 
non-$B$ background.
}
\label{fig:mkpi_signal}
\end{figure}

The $M(K\pi)$ distribution for events within $\pm 0.03$~GeV of the
peak at 4.43~GeV is shown in Fig.~\ref{fig:mkpi_signal}.  Here the
$K^*$~veto, which excludes the regions indicated by the
double-sided arrows in the figure, has been removed.  
The shaded histogram is the
scaled $\DE$ sideband data. Aside from the $\kstr$ resonance events,
which are removed by the $K^*$~veto, no dramatic features are evident.

We considered the possibility that
interference between $S$-, $P$- and
$D$-waves in the $K\pi$ system might produce a structure
similar to that which is observed. 
(There are $F$-wave and higher $K\pi$ resonances listed in
the PDG tables, however, even the lowest mass $F$-wave entry, 
the $K^*_3(1780)$, is not kinematically 
accessible in $B\rt K\pi\psip$ decay.)
We find that with only
these three partial waves, it is not possible
to produce a $\pi\psip$ invariant mass  peak near 4.43~GeV
that is as narrow as the one we see without other,
even more dramatic, accompanying structures.

We applied the same analysis to large MC samples of generic
$B$ meson decays and found no evidence of peaking in the
$\pi\psip$ invariant mass distribution.


\label{sec:prodbr_calc} 
The product branching fraction is determined 
using MC-computed acceptance values and world average 
values for $\psip$ and $\jp$ branching fractions~\cite{PDG}.
 For this calculation, we only
use the signal yield from the $B^0\rt K^{\mp}\pi^{\pm}\psip$ 
decay sample.  
The resulting product
branching fraction is 
\begin{eqnarray}
\nonumber
  {\mathcal B}(\bar{B^0}\rt K^- Z^+(4430))\times {\mathcal B}(Z^+(4430)\rt 
\pi^+\psip) &~~~&\\  = (4.1 \pm 1.0 \pm 1.4)\times 10^{-5},&~~~&
\end{eqnarray}
where $Z^+(4430)$ is used to denote the observed structure,
the first error is statistical and the second error
is systematic (discussed below).


The values of the $\psip$ mass determined using events in the data
agree within $\pm 1$~MeV with the world average value for both the
$\psip\rt\pipi\jp$ and $\psip\rt\leplep$ decay modes.   We find
less than 1~MeV variation in the peak mass value for different
fitting functions.  Fits that include possible interference between the
BW signal and the non-resonant $\pi\psip$ continuum
produce at most a 1.2~MeV shift in the fitted mass  
value.  We assign
a $\pm 2$~MeV systematic error to the mass determination.

The systematic uncertainty on the width is mostly due to the
uncertain effects of background fluctuations feeding into the fitted 
signal. We estimate the level of this effect from the range in width 
values determined  from different subsets of the data to be 
$+29$~MeV and $-10$~MeV.
Changes in the parameterization of
$f_{\rm cont}(M)$ and variations in the range
of $M(\pi\psip)$ values included in the fit produce $\pm 8$~MeV
changes in the width; using different BW forms produce $\pm 4$~MeV
width changes.  Adding these sources in quadrature results in
a total systematic error on the width of $^{+30}_{-13}$~MeV. 

The largest systematic error on the product branching fraction
measurement is due to the correlation between the fitted signal yield
and the peak width.  A $+ 30 $~MeV ($ -13$~MeV)
change in the width produces a $ + 30\%$ ($-17\% $)
change in signal yield.  Changes in the
parameterization of $f_{\rm cont}(M)$ produce
 $+ 11\% $ and $- 18\% $ variations in
signal yield. 
Other systematic errors are smaller.  These
include: possible interference with the $\pi\psip$ continuum;
the choice of BW signal function; uncertainties in the
acceptance calculation; uncertainties
in the tracking and particle identification efficiencies;
errors on the world-average $\psip$ decay
branching fractions; MC statistics; and the error
on the number of $B\bar{B}$ mesons in the sample.
Combining these errors in quadrature gives a systematic error
on the product branching fraction of 35\%.


In summary, a
study of $B\rt K\pi^+\psip$ decays reveals a 
peak in the $\pi^+\psip$ invariant mass spectrum at 
$M = (4433 \pm 4 {\rm (stat)} \pm 2 {\rm (syst)})~{\rm MeV}.$
The measured width,
$\Gamma= (45^{+18}_{-13} {\rm (stat)} ^{+30}_{-13}
{\rm (syst)})~{\rm MeV},$
is too  narrow to be caused by interference effects in
the $K\pi$ channel.   
The statistical significance of the
observed peak is $6.5\sigma$.



There have been a number of anomalous charmonium-like meson
candidates reported in the 
literature~~\cite{skchoi_etac2s,uehara_chic2,skchoi_x3872,aubert_y4260,
kabe_x3940,skchoi_y3940,aubert_y4325,yuan_y4096,yuan_y4660}.  
The structure reported here
is unique in that it is the first candidate to have a non-zero
electric charge.


We thank the KEKB group for excellent operation of the
accelerator, the KEK cryogenics group for efficient solenoid
operations, and the KEK computer group and
the NII for valuable computing and Super-SINET network
support.  We acknowledge support from MEXT and JSPS (Japan);
ARC and DEST (Australia); NSFC and KIP of CAS (China); 
DST (India); MOEHRD, KOSEF and KRF (Korea); 
KBN (Poland); MES and RFAAE (Russia); ARRS (Slovenia); SNSF (Switzerland); 
NSC and MOE (Taiwan); and DOE (USA).


%


\begin{thebibliography}{99}
\bibitem{conj}
In this letter the inclusion
of the charge conjugate mode is always implied.  The $\psi'$ is sometimes
denoted as $\psi(2S).$
\bibitem{thetaB} The $\theta_B$ distribution has a $\sin^2\theta_B$
dependence for $\ee\rt B\bar{B}$ and is flat for
continuum background events.
\end{thebibliography}
\end{document}